\documentclass[prb,aps,twocolumn,showpacs,nofootinbib]{revtex4}

\usepackage{graphicx}
\usepackage{amsmath}
\usepackage{amsfonts}
\usepackage{amssymb}
\usepackage{dcolumn} % align table columns on decimal point
\usepackage{bm} % bold math
\usepackage{color}

\usepackage{latexsym}

\begin{document}
\title{Implications of resonant inelastic X-ray scattering data for theoretical models of cuprates} 
%Enigma of resonant inelastic X-ray scattering  from cuprates}
%Is $t-t^{\prime}-t^{\prime\prime}-J$ model sufficient to describe magnetic excitations in cuprates?}

\author{Wei Chen$^{1}$ and Oleg P. Sushkov$^{2}$}
\affiliation{
$^1$Max Planck Institute for Solid State Research, Heisenbergstrasse 1, D-70569 Stuttgart, Germany \\
$^2$School of Physics, University of New South Wales, 
Sydney 2052, Australia }

\date{\today}

\begin{abstract}
There are two commonly discussed points of view in theoretical description of cuprate superconductors,
(i) Cuprates can be described by the modified $t-J$ model.
(ii) Overdoped cuprates are close to the regime of normal Fermi liquid (NFL).
We argue that recent resonant inelastic X-ray scattering data challenge  both points.
While the modified $t-J$ model  describes well the strongly underdoped regime, it fails to describe 
high energy magnetic 
excitations when  approaching optimal doping. This probably indicates failure of the Zhang-Rice singlet picture.
In the overdoped regime the momentum-integrated spin structure factor $S(\omega)$ has the same intensity and
energy distribution as that in an undoped parent compound. This implies that the entire spin spectral sum rule
is saturated at $\omega \approx 2J$, while in a NFL the spectral weight should saturate only
at the total bandwidth which is much larger than $2J$.
\end{abstract}

\date{\today}


\maketitle

\section{Introduction}

Parent compounds for cuprate superconductors are antiferromagnetic (AF) charge transfer insulators (CTI) with localized 
spins~\cite{CT,Chen91}.
Magnetic excitations in the parent compounds are usual Heisenberg model spin waves
with maximum energy $\omega_{max}\approx 2J \approx 300$meV~\cite{expMott}.
Conductivity and superconductivity arises when cuprates are doped with mobile charge carriers.
In this paper we discuss only the hole doping. 
It is generally believed that magnetic excitations play a crucial role in the superconducting 
pairing 
mechanism. Magnetic structure of cuprates and magnetic excitations dramatically evolve 
with doping. Magnetic structure has been studied in elastic, quasielastic, and  inelastic
neutron scattering, as well as in $\mu$SR~\cite{Yamada98,Hinkov07,Stock08,haug09,haug10,connery10}.
The three dimensional (3D) long range AF order is preserved at a 
very low doping up to some doping level $p_1$. The value of $p_1$ is about few per cent and the value depends
on a particular compound and a particular way of doping.
An incommensurate spin structure starts to develop at $p > p_1$. The incommensurability
destroys the 3D coupling between CuO$_2$ planes and  hence the magnetic structure becomes quasistatic in 
the incommensurate phase.
The quasistatic magnetic order exists up to doping $p_2$ which is a magnetic quantum critical point (QCP).
The QCP is a generic property and position of the QCP is approximately the same for all
cuprates, $p_2 \approx 10\%$.
At $p > p_2$ the incommensurate magnetic structure becomes fully dynamic.
The evolution of the magnetic structure is perfectly consistent with predictions of the $t-t'-t''-J$ 
model~\cite{luscher07,milstein08,sushkov09}. The incommensurate magnetic structure is a spin spiral.
The evolution of the magnetic  structure is accompanied by redistribution of the magnetic spectral
weight. The static magnetic response is transferred to the ``resonant'' energy, the position of the
narrow neck in the hourglass dispersion~\cite{milstein08}.
The described evolution studied in neutron scattering is relevant to the static magnetic structure and
to the low energy, $\epsilon < 100$meV, magnetic excitations with wave vector close to 
$(\pi,\pi)$. 
Magnetic excitations with energy higher than 100meV in doped cuprates are  difficult to assess with neutrons.

Recent development in resonant inelastic x-ray scattering
(RIXS)~\cite{Braicovich09,Braicovich10,LeTacon11,Bisogni12,Bisogni12_2,Dean12,Dean13,LeTacon13} 
has made it possible 
to probe high energy magnetic excitations in cuprates, complementary to the low energy regime studied by 
neutrons.  To avoid ambiguity in describing data we do not use terms magnon or paramagnon, 
instead we use the term ``magnetic excitation''. The excitation can contain one, two, or more  
magnons (paramagnons).  According to the RIXS data the energy width of the magnetic excitation is rather large, 
$\Gamma \approx 200$meV. The large width itself is not too surprising. 
In contrast, the following two points concluded by RIXS are far more surprising:
(i) At a given momentum $q$ the magnetic response is a broad peak positioned at some energy.
%$\epsilon_{max}({\bm q})$.
The position of the peak is independent of doping. In underdoped, optimally doped, and even overdoped  cuprates
 the position is the same as that in undoped CTI~\cite{LeTacon11,LeTacon13}.
(ii) The energy-integrated spectral weight at a given momentum
is doping-independent. Again,
in underdoped, optimally doped, and even overdoped  cuprates
the spectral weight is the same as that in undoped CTI~\cite{LeTacon11,LeTacon13}.
Thus, while the low energy, $\epsilon < 100$meV, magnetic response evolves dramatically with doping,
the high energy response, $100\mbox{meV} < \epsilon < 2J\approx 300$meV practically
does not evolve with doping (besides the line broadening).

The $t-J$ model was suggested phenomenologically at the very early stage of cuprate physics~\cite{Anderson87}.
It became clear very soon that one needs to slightly extend the model by introducing additional
hopping matrix elements $t'$, $t''$.
% to describe the charge carrier dispersion.
The additional matrix elements are qualitatively important, for example they destroy the electron-hole
symmetry. The asymmetry explains a very significant difference between the hole and the electron doping.
Dynamics of CuO$_2$ planes in cuprates is determined by 3d electrons of copper and 2p electrons of oxygen.
In the case of hole doping very few holes go to 3d states
of Cu. Even in overdoped samples the concentration of 3d holes differs from that in the parent CTI
insulator only by a few per cent~\cite{haase04}. Doped holes go predominantly to 2p states of oxygen. This
 is a qualitative difference of a doped CTI from a doped Mott insulator.
In this  situation the $t-t'-t''-J$ model can be justified only by  the 
Zhang-Rice singlet picture~\cite{Zhang88}.
The picture seems well justified when the hole momentum  is close to $(\pm \pi/2,\pm \pi/2)$, however, away
from this point the picture is questionable as noted already in the original 
paper~\cite{Zhang88}. This implies that the justification of the $t-t'-t''-J$ model
is getting more and more questionable when the doping is  increasing, see also Ref.~\onlinecite{Lau11}.

In the present paper we address theoretically two issues (i) dependence  of high energy magnetic 
excitations on doping, (ii) momentum integrated spin some rule.
Concerning the first issue we show that the $t-t'-t''-J$ model predicts a significant softening of the high energy
magnetic response with doping. The softening is inconsistent with RIXS data.
With respect to the second issue we argue that the RIXS data are inconsistent with the picture of almost
normal Fermi liquid in heavily overdoped cuprates. Here, in essence we reiterate the claim already made in the 
experimental  paper~\cite{LeTacon13}. We just use a different language to make the same point.

The structure of the paper is the following. In Sec. II  we calculate softening of high energy magnetic 
response. This section is relevant to the underdoped regime because the applied theoretical technique
is not valid above optimal doping.
In Sec. III we consider the exact spin sum rule. The sum rule is valid for any doping, however the most important
conclusion comes for the heavily overdoped regime. So this Section is mainly aimed at overdoped cuprates.
Our conclusions are summarized in Sec. IV.

\section{Softening of high energy magnetic response with doping}

We have already pointed out that the low energy magnetic response of cuprates evolves dramatically
with doping. On the theoretical side the response is well described by the $t-t'-t''-J$ 
model. Calculation of the low energy response within the model is a highly nontrivial theoretical problem.
The only controlled approach to this problem is the chiral perturbation theory used in 
Refs.~\onlinecite{luscher07,milstein08,sushkov09}.
The perturbation theory uses the parent CTI as zero approximation and then allows to derive various physical 
properties as expansions in powers of doping $p$. The theory allows to calculate leading in $p$ 
terms as well as first corrections. The expansion is in powers of $\sqrt{p}$, so the expansion is 
nonanalytic.
We stress that usually only the first correction can be calculated in a fully controlled way.
For example the wave vector of the spin spiral which arises in the model scales as $p$ and 
the first correction to the wave vector $\propto p^{3/2}$ is exactly zero.
The static on-site magnetization behaves as $m=0.6\mu_B-a\sqrt{p}+bp$ where the coefficient $a$ has 
been reliably calculated and the coefficient $b$ has been estimated~\cite{milstein08}. 
The position of the narrow neck in the hourglass dispersion scales as $E_{res}\propto p^{3/2}$.
The presented scalings are valid for a single layer cuprate, for a double layer 
scalings are somewhat different\cite{sushkov09a,sushkov11}.

High energy properties of the $t-t'-t''-J$ model are much simpler.
The chiral perturbation theory implies that the short range correlations are
independent of doping. This seems consistent with RIXS data and this implies
that the high energy properties are unchanged at least in the first order of the
chiral perturbation theory, i. e. the correction $\propto \sqrt{p}$ is zero. 
``High energy'' here means that $\omega >> E_{res}$.
We remind that near optimal doping $E_{res} \approx 40-50$meV~\cite{RM91,Bourges05}.
High energy properties still can change in the subleading order.
So a variation  proportional to $ (\sqrt{p})^2 =p$ is possible.
In this section we calculate this variation.
The results of the present section are valid up to optimal doping, $p\lesssim 0.15$. 
Below we explain why the present section calculation is not justified at  $p > 0.15$.

The magnetic background fluctuates with typical frequencies $\omega \sim E_{res}$.
These quantum fluctuations  lead to the dramatic change of the static and the low energy response.
However, the fluctuations are irrelevant for $\omega  \gg E_{res}$, the high energy
excitation ``sees'' a snapshot of the magnetic background and the snapshot is the usual
collinear AF. So, here we take the simple AF background and calculate the magnetic response
using  self-consistent Born approximation (SCBA). The applied techniques are  similar
to that used long time ago in Refs.~\cite{Igarashi92,Sherman94,Plakida94,Kyung96}.
However, now we have a better level of understanding.
For instance, we know that the approach cannot capture the 
static and the low energy sector, and we also know about the qualitative importance of $t'$ and $t''$.

\subsection{Hole spectral function at finite doping}
SCBA has been widely adopted to study the hole dynamics in the presence of AF magnetic order 
for either single hole\cite{Kane89,Martinez91,Liu92,Ramsak92,Sushkov97} or at finite
doping\cite{Igarashi92,Sherman94,Plakida94,Kyung96}. 
In this subsection we remind the major steps of SCBA.
The Hamiltonian of the $t-t'-t''-J$ model reads
\begin{eqnarray}
H_{t}+H_{t^{\prime},t^{\prime\prime}}+H_{J}&=&\sum_{i,j,\sigma}-t_{ij}c_{i\sigma}^{\dag}c_{j\sigma}
+J\sum_{\langle ij\rangle}{\bf S}_{i}\cdot{\bf S}_{j}
\label{single_layer_H}
\end{eqnarray}
where $c_{i\sigma}$ is the electron annihilation operator of spin $\sigma$ at site $i$, 
$t_{ij}\in\left\{t,t^{\prime},t^{\prime\prime}\right\}$ is the nearest, next-nearest, and next-next-nearest 
neighbor hopping, respectively. On the top of the Hamiltonian (\ref{single_layer_H}) one has to add a
no double electron occupancy constraint. We set energy unit $J\approx 140$meV$\rightarrow 1$. In this section we 
choose $t=3.1$, $t^{\prime}=-0.5$, $t^{\prime\prime}=0.4$ according  to fitting ARPES in undoped parent compound~\cite{Chen11}.
In the next subsection we also calculate magnon spectral function in the pure $t-J$ model 
($t^{\prime}=t^{\prime\prime}=0$) to demonstrate the generic feature of this type of models.

AF background implies two sublattices, 'up' and 'down', and this allows to introduce two 
spin-wave excitations, $a_{i}^{\dag}\in\uparrow$ and $b_{j}^{\dag}\in\downarrow$. 
Using Fourier and Bogoliubov transformation
\begin{eqnarray}
a_{{\bf q}}=u_{{\bf q}}\alpha_{{\bf q}}+v_{{\bf q}}\beta_{{\bf -q}}^{\dag}
\;,\;\;\;b_{{\bf -q}}=v_{{\bf q}}\alpha_{{\bf q}}^{\dag}+u_{{\bf q}}\beta_{{\bf -q}}
\label{Bogoliubov_trans}
\end{eqnarray}
one can diagonalize the $H_J$ part of Hamiltonian~(\ref{single_layer_H})
\begin{eqnarray}
H_{J}&=&\sum_{\bf q}\left(\alpha_{{\bf q}}^{\dag}\alpha_{{\bf q}}+\beta_{{\bf q}}^{\dag}\beta_{{\bf q}}\right)\omega_{{\bf q}}\;,
\nonumber \\
\omega_{{\bf q}}&=&2\sqrt{1-\gamma_{\bf q}^{2}}\;,
\label{single_layer_dispersion}
\end{eqnarray}
where the Bogoliubov coefficients are
\begin{eqnarray}
\label{bog}
u_{{\bf q}}=\sqrt{\frac{1}{\omega_{{\bf q}}}+\frac{1}{2}}
\;,\;\;\;v_{{\bf q}}=-{\rm sign}({\gamma_{\bf q}})\sqrt{\frac{1}{\omega_{{\bf q}}}-\frac{1}{2}}\;.
\end{eqnarray}
The corresponding bare magnon Green's function is
\begin{eqnarray}
&&D^{0}(\omega,{\bf q})=-i\int_{-\infty}^{\infty}\langle T\alpha_{\bf q}(t)\alpha_{\bf q}^{\dag}(0)\rangle e^{i\omega t}dt
 \\
&&=-i\int_{-\infty}^{\infty}\langle T\beta_{\bf -q}^{\dag}(t)\beta_{\bf -q}(0)\rangle e^{-i\omega t}dt
=\frac{1}{\omega-\omega_{\bf q}+i\eta} . \nonumber
\label{bare_magnon_diagonal}
\end{eqnarray}
The hole operators with pseudospin up and down are defined in different sublattices\cite{Sushkov97}
\begin{eqnarray}
&&d_{{\bf k}\uparrow}=\sqrt{\frac{2}{N}}\sum_{j}c_{j\downarrow}^{\dag}e^{-i{\bf k\cdot r}_{j}}
\nonumber \\
&&d_{{\bf k}\downarrow}=\sqrt{\frac{2}{N}}\sum_{i}c_{i\uparrow}^{\dag}e^{-i{\bf k\cdot r}_{i}}
\end{eqnarray}
where $N$ is the number of lattice sites.
The bare hole dispersion is then given by $H_{t^{\prime},t^{\prime\prime}}$
\begin{eqnarray}
&&H_{t^{\prime},t^{\prime\prime}}=\sum_{{\bf k},\sigma}\epsilon_{{\bf k}}^{(0)}d_{{\bf k}\sigma}^{\dag}d_{{\bf k}\sigma}\;,
\nonumber \\
&&\epsilon_{{\bf k}}^{(0)}=4t^{\prime}\cos k_{x}\cos k_{y}+2t^{\prime\prime}\left(\cos 2k_{x}+\cos 2k_{y}\right) .
\end{eqnarray}
The hole-magnon vertex comes from $H_{t}$, see e.g. Ref.~\cite{Sushkov97}
\begin{eqnarray}
&&H_{t}=\sum_{{\bf k,q}}g_{\bf k,q}\left(d_{{\bf k+q}\downarrow}^{\dag}d_{{\bf k}\uparrow}\alpha_{{\bf q}}+d_{{\bf k+q}\uparrow}^{\dag}d_{{\bf k}\downarrow}\beta_{{\bf q}}\right)+h.c.
\nonumber \\
&&g_{{\bf k,q}}=4t\sqrt{\frac{2}{N}}\left(\gamma_{\bf k}u_{{\bf q}}+\gamma_{\bf k+q}v_{{\bf q}}\right)
\label{hole_magnon_vertices}
\end{eqnarray}

\begin{figure}[ht]
\begin{center}
\includegraphics[clip=true,width=0.8\columnwidth]{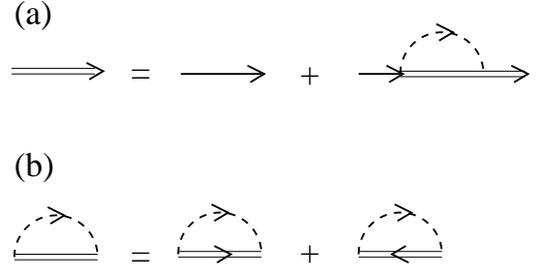}
\caption{(a)Dyson's equation of hole Green's function described by Eq.~(\ref{Gd_SCBA}). (b)Retarded and advanced part of hole self-energy, Eq.~(\ref{hole_self_energy}). 
 }
\label{SE}
\end{center}
\end{figure}

SCBA is equivalent to summation of magnon rainbow diagrams for the hole Green's function
$G_{d}$. The summation is equivalent to solution of Dyson equation
\begin{eqnarray}
&&G_{d}(\epsilon,{\bf k})=\left[\epsilon-\epsilon_{\bf k}^{0}-\Sigma(\epsilon,{\bf k})+ i\eta \ \mbox{sign}(\epsilon-\mu)
\right]^{-1}
\label{Gd_SCBA}
\end{eqnarray} 
shown diagrammatically in Fig.~\ref{SE} (a), $\Sigma$ is the self-energy.
Note that at a finite doping one must use Feynman Green's function, therefore the imaginary shift of 
the denominator, $\pm i \eta$, depends on energy and  chemical potential $\mu$.
This leads to complications in numerical solution of Eq. (\ref{Gd_SCBA}).
To overcome these complications we utilize spectral representation for the Feynman Green's function 
\begin{eqnarray}
&&G_{d}(\epsilon,{\bf k})=-i\int_{-\infty}^{\infty}\langle Td_{{\bf k}\sigma}(t)d_{{\bf k}\sigma}^{\dag}(0)\rangle e^{i\epsilon t}dt
\nonumber \\
&=&\int_{\mu}^{\infty}dx\frac{ A(x,{\bf k})}{\epsilon-x+i\eta}+\int_{-\infty}^{\mu}dx\frac{ B(x,{\bf k})}{\epsilon-x-i\eta}\;,
\label{Gd_spectral_representation}
\end{eqnarray}
and solve Eq. (\ref{Gd_SCBA}) with respect to spectral densities $A(x,{\bf k})$ and $B(x,{\bf k})$.
The self-energy is determined by diagrams in Fig.~\ref{SE} (b)
\begin{eqnarray}
&&\Sigma(\epsilon,{\bf k})
=\int\frac{d^{2}{\bf q}}{(2\pi)^{2}}g_{\bf k-q,q}^{2}
\int_{\mu}^{\infty}dx\frac{A(x,{\bf k-q})}{\epsilon-\omega_{\bf q}-x+i\eta}
\nonumber \\
&&+\int\frac{d^{2}{\bf q}}{(2\pi)^{2}}g_{\bf k,-q}^{2}\int_{-\infty}^{\mu}dx\frac{B(x,{\bf k-q})}{\epsilon+\omega_{\bf q}-x-i\eta}\;.
\label{hole_self_energy}
\end{eqnarray}
The hole density is given by the negative frequency part of the spectral function.
\begin{eqnarray}
p=2\int\frac{d^{2}{\bf k}}{(2\pi)^{2}}\int_{-\infty}^{\mu}dx B(x,{\bf k})\;.
\label{hole_density}
\end{eqnarray}
All momentum integration are limited inside magnetic Brillouin zone (MBZ).
Note that in (\ref{hole_self_energy}) we use bare magnons described by Eq. (\ref{bare_magnon_diagonal}).

\begin{figure}[ht]
\begin{center}
\includegraphics[clip=true,width=0.95\columnwidth]{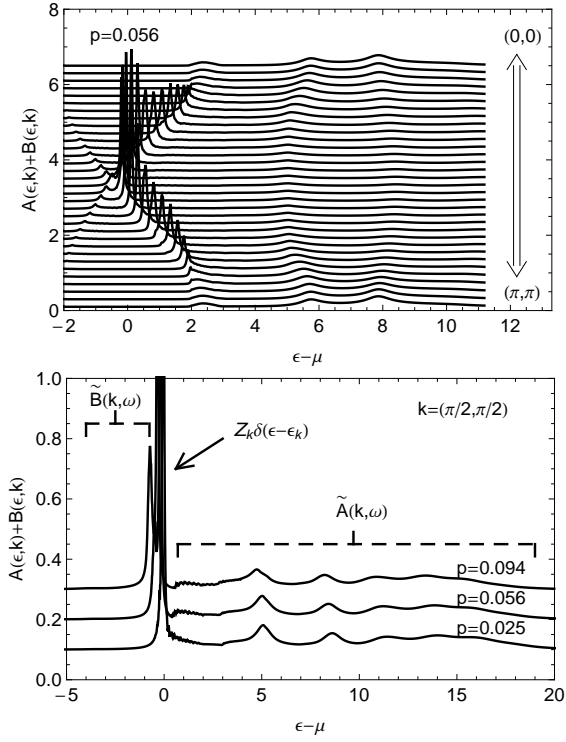}
\caption{Top panel: Hole spectral function along nodal direction at doping $p=0.056$.
Bottom panel: Hole spectral function at nodal point ${\bf k}=(\pi/2,\pi/2)$ at different doping levels.}
\label{fig:hole_spectral_function}
\end{center}
\end{figure}
Spectral function obtained  by numerical solution of Dyson Eq. is shown in Fig.~\ref{fig:hole_spectral_function}.
We plot $A(\epsilon,{\bm k})$ when $\epsilon-\mu > 0$ and $B(\epsilon,{\bm k})$ when $\epsilon-\mu <0$.
The top panel in Fig.~\ref{fig:hole_spectral_function} shows the spectral function along
the nodal direction for doping $p=0.056$.
Note that the spectral function is identical inside and outside of MBZ.
This is not the spectral function measured in angle resolved photoemission (ARPES).
To recover the ARPES spectral function, one needs to take into account additional diagrams~\cite{Sushkov97}.
The ARPES spectral function is highly asymmetric with respect to MBZ with only a tiny intensity
outside of MBZ~\cite{Chen11}.
The bottom panel in Fig.~\ref{fig:hole_spectral_function} shows the spectral function for
${\bm k}=(\pi/2,\pi/2)$ and for three different values of doping.
As usually spectral functions contain coherent quasiparticle peaks, 
and a large incoherent background that extends over a wide energy range of the
free hole band width, $\Delta E \approx 8t\approx 24$. 
So, we can represent spectral functions as
\begin{eqnarray}
&&A(\epsilon,{\bf k})=Z_{{\bf k}}\delta\left(\epsilon-\epsilon_{{\bf k}}\right)+\tilde{A}(\epsilon,{\bf k})
\nonumber \\
&&B(\epsilon,{\bf k})=Z_{{\bf k}}\delta\left(\epsilon-\epsilon_{{\bf k}}\right)+\tilde{B}(\epsilon,{\bf k})
\label{QP_approximation}
\end{eqnarray}
Schematics of this separation are shown in the bottom panel of Fig.~\ref{fig:hole_spectral_function}.
The quasiparticle residue at ${\bf k}=(\pi/2,\pi/2)$ is about $Z\approx 0.35$. 
The residue gradually decreases as moving away from ${\bf k}=(\pi/2,\pi/2)$
practically vanishing at top of the band.
The hole dispersion, identified by position of the quasiparticle peak, clearly shows a hole 
pocket centering at ${\bf k}=(\pi/2,\pi/2)$. 
Similar to the investigation in double layer compound,\cite{Chen11} 
the hole dispersion is practically rigid against hole doping, i.e. the shape of the dispersion 
is roughly unchanged at small doping, despite a minor correction on ellipticity of hole pockets
(doping makes the pockets more elliptic).

The Fermi energy scales linearly with doping. According to Fig.~\ref{fig:hole_spectral_function}
the Fermi energy at $p=0.094$ is $\epsilon_F\approx 0.6J\approx 90$meV.
We already pointed out that assuming AF ordering we make a snapshot of
fluctuating magnetic background. The background fluctuates with typical frequency
about $E_{res}\propto p^{3/2}$, see Ref.~\cite{milstein08}
For validity of the snapshot approach the Fermi energy,
$\epsilon_F \propto p$, must be much larger than $E_{res}$, $E_{res} \ll \epsilon_F$.
The inequality is violated at about optimal doping. This is why our snapshot approach
is justified only at $p < 0.15$.

\subsection{Magnon softening at finite doping}
\begin{figure}[ht]
\begin{center}
\includegraphics[clip=true,width=0.95\columnwidth]{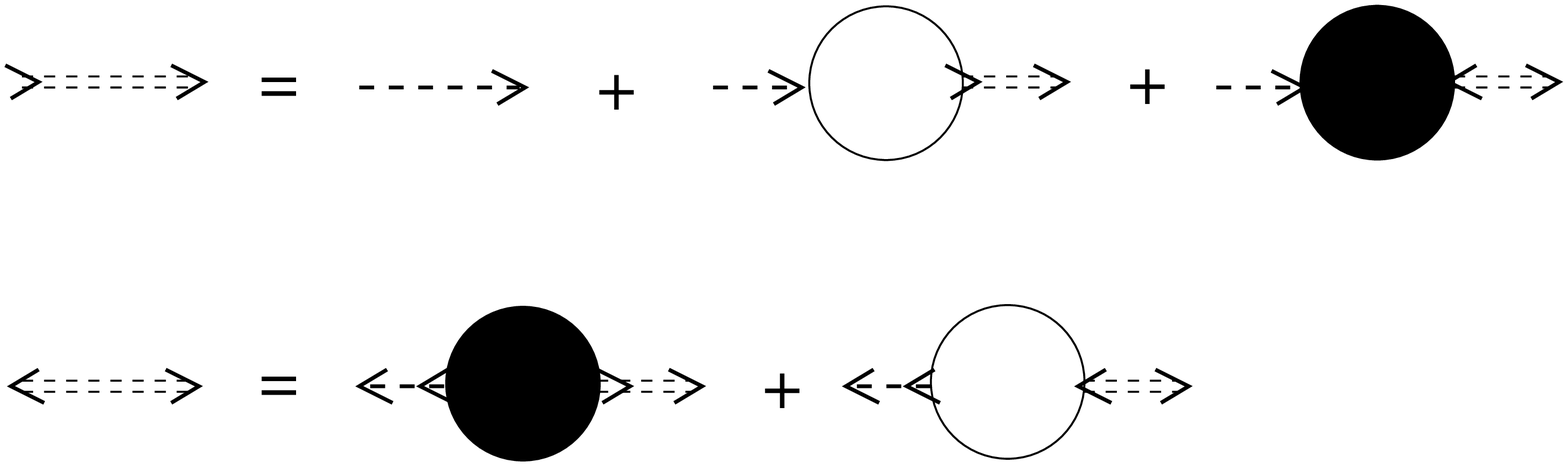}
\caption{Dyson's equations of magnon Green's function, Eq.~(\ref{Dyson_magnon}). 
The unfilled circle shows the normal polarization operator and the full 
circle shows the anomalous polarization operator.
}
\label{pol}
\end{center}
\end{figure}

To address the renormalization of magnon due to $G_{d}$, we define the following matrix elements for dressed magnon Green's functions\cite{Igarashi92} 
\begin{eqnarray}
D_{\alpha\alpha}(q)&=&-i\int_{-\infty}^{\infty}\langle T\alpha_{{\bf q}}(t)\alpha_{{\bf q}}^{\dag}(0)\rangle e^{i\omega t}dt=D(q)
\nonumber \\
D_{\beta\beta}(q)&=&-i\int_{-\infty}^{\infty}\langle T\beta_{{\bf -q}}^{\dag}(t)\beta_{{\bf -q}}(0)\rangle e^{i\omega t}dt=D(-q)
\nonumber \\
D_{\alpha\beta}(q)&=&-i\int_{-\infty}^{\infty}\langle T\alpha_{{\bf q}}(t)\beta_{{\bf -q}}(0)\rangle e^{i\omega t}dt=\overline{D}(q)
\label{magnon_GF_definition}
\end{eqnarray}
where $q=(\omega,{\bf q})$. Dyson's equation for $D(q)$ and $\overline{D}(q)$ is\cite{Igarashi92,Khaliullin93}
\begin{eqnarray}
D(q)&=&D^{0}(q)+D^{0}(q)\Pi_{11}(q)D(q)
+D^{0}(q)\Pi_{02}(q)\overline{D}(q)
\nonumber \\
\overline{D}(q)&=&D^{0}(-q)\Pi_{20}(q)D(q)
+D^{0}(-q)\Pi_{11}(-q)\overline{D}(q)
\nonumber \\
&&
\label{Dyson_magnon}
\end{eqnarray}
These equations are graphically presented in Fig.~\ref{pol}. Here $\Pi_{11}$ and $\Pi_{02}=\Pi_{20}$ 
are the normal and the anomalous polarization operators respectively.
Solution to Eq. (\ref{Dyson_magnon}) is
\begin{eqnarray}
D(q)&=&\frac{D^{0}(-q)^{-1}-\Pi_{11}(-q)}{\lambda(q)}\;,\;\;\;
\overline{D}(q)=\frac{\Pi_{20}(q)}{\lambda(q)}\;,
\nonumber \\
\lambda(q)&=&\left[D^{0}(q)^{-1}-\Pi_{11}(q)\right]
\left[D^{0}(-q)^{-1}-\Pi_{11}(-q)\right]
\nonumber \\
&&-\Pi_{02}(q)\Pi_{20}(q)
\label{magnon_SCBA_solution}
\end{eqnarray}
The normal and the anomalous  polarization operators
are calculated by the spectral function of holes 
\begin{eqnarray}
\Pi_{11}(q)&=&\int \frac{d^{2}{\bf k}}{(2\pi)^{2}}g^{2}_{{\bf k-q,q}}P(\omega,{\bm k},{\bm q})
\nonumber \\
\Pi_{02}(q)&=&\int \frac{d^{2}{\bf k}}{(2\pi)^{2}}g_{{\bf k-q,q}}g_{{\bf k,-q}}P(\omega,{\bm k},{\bm q})\;,
\nonumber \\
P(\omega,{\bm k},{\bm q})&=&\int_{-\infty}^{\mu}d\epsilon\int_{\mu}^{\infty}d\epsilon^{\prime}
\left[\frac{A(\epsilon^{\prime},{\bf k})B(\epsilon,{\bf k-q})}{\omega+\epsilon-\epsilon^{\prime}+i\eta}\right.
\nonumber \\
&&\;\;\;-\left.\frac{B(\epsilon,{\bf k})A(\epsilon^{\prime},{\bf k-q})}{\omega+\epsilon^{\prime}-\epsilon-i\eta}\right] \ .
\label{polarization_1st_expression}
\end{eqnarray}
%and $\Pi_{02}(q)=\Pi_{20}(q)$. 

Integrals in Eq. (\ref{polarization_1st_expression}) are pretty singular having in mind that 
the spectral functions contain coherent $\delta$-function contributions.
Therefore, for accurate numerical integration we split the coherent and incoherent dynamics.
Practically this means that the integrand in (\ref{polarization_1st_expression}) is 
split in four parts which we calculate separately.
\begin{eqnarray}
P=P^{ZZ}&+&P^{Z\tilde{A}}+P^{Z\tilde{B}}+P^{\tilde{A}\tilde{B}}\;,
\nonumber \\
P^{ZZ}(k,q)&=&\frac{Z_{{\bf k}}Z_{{\bf k-q}}\Theta(\epsilon_{\bf k}-\mu)\Theta(-\epsilon_{\bf k-q}+\mu)}{\omega+\epsilon_{{\bf k-q}}-\epsilon_{{\bf k}}+i\eta}
\nonumber \\
&&\;\;\;-\frac{Z_{{\bf k}}Z_{{\bf k-q}}\Theta(-\epsilon_{\bf k}+\mu)\Theta(\epsilon_{\bf k-q}-\mu)}{\omega+\epsilon_{{\bf k-q}}-\epsilon_{{\bf k}}-i\eta}\;,
\nonumber \\
P^{Z\tilde{A}}(k,q)&=&\int_{\mu}^{\infty}d\epsilon^{\prime}\frac{Z_{{\bf k-q}}\Theta(-\epsilon_{\bf k-q}+\mu)\tilde{A}(\epsilon^{\prime},{\bf k})}{\omega+\epsilon_{{\bf k-q}}-\epsilon^{\prime}+i\eta}
\nonumber \\
&&\;\;\;-\int_{\mu}^{\infty}d\epsilon^{\prime}\frac{Z_{k}\Theta(-\epsilon_{\bf k}+\mu)\tilde{A}(\epsilon^{\prime},{\bf k-q})}{\omega+\epsilon^{\prime}-\epsilon_{\bf k}-i\eta}\;,
\nonumber \\
P^{Z\tilde{B}}(k,q)&=&\int_{-\infty}^{\mu}d\epsilon\frac{Z_{{\bf k}}\Theta(\epsilon_{\bf k}-\mu)\tilde{B}(\epsilon,{\bf k-q})}{\omega+\epsilon-\epsilon_{{\bf k}}+i\eta}
\nonumber \\
&&\;\;\;-\int_{-\infty}^{\mu}d\epsilon\frac{Z_{{\bf k-q}}\Theta(\epsilon_{\bf k-q}-\mu)\tilde{B}(\epsilon,{\bf k})}{\omega+\epsilon_{\bf k-q}-\epsilon-i\eta}\;,
\nonumber \\
P^{\tilde{A}\tilde{B}}(k,q)&=&\int_{-\infty}^{\mu}d\epsilon\int_{\mu}^{\infty}d\epsilon^{\prime}
\left[\frac{\tilde{A}(\epsilon^{\prime},{\bf k})\tilde{B}(\epsilon,{\bf k-q})}{\omega+\epsilon-\epsilon^{\prime}+i\eta}\right.
\nonumber \\
&&\;\;\;\left.-\frac{\tilde{B}(\epsilon,{\bf k})\tilde{A}(\epsilon^{\prime},{\bf k-q})}{\omega+\epsilon^{\prime}-\epsilon-i\eta}\right]\;.
\label{P_ZZ_ZA_ZB_AB}
\end{eqnarray}
\begin{figure}[ht]
\begin{center}
\includegraphics[clip=true,width=0.85\columnwidth]{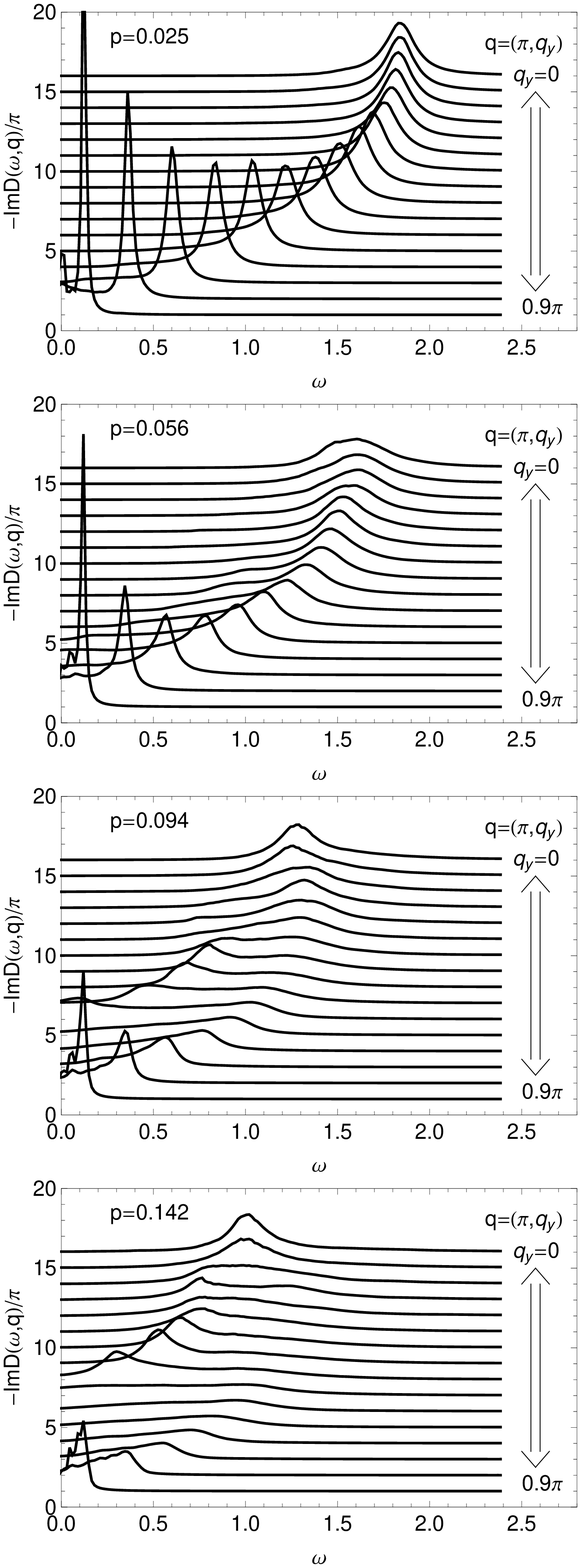}
\caption{Imaginary part of Magnon Green's function obtained by numerical method, Eqs. (\ref{QP_approximation}) and (\ref{P_ZZ_ZA_ZB_AB}), at different doping levels for the $t-t^{\prime}-t^{\prime\prime}-J$ model with $t^{\prime}=-0.5$ and $t^{\prime\prime}=0.4$. } 
\label{fig:ImD_ZZ_ZA_ZB_AB_scandoping}
\end{center}
\end{figure}
\begin{figure}[ht]
\begin{center} 
\includegraphics[clip=true,width=0.85\columnwidth]{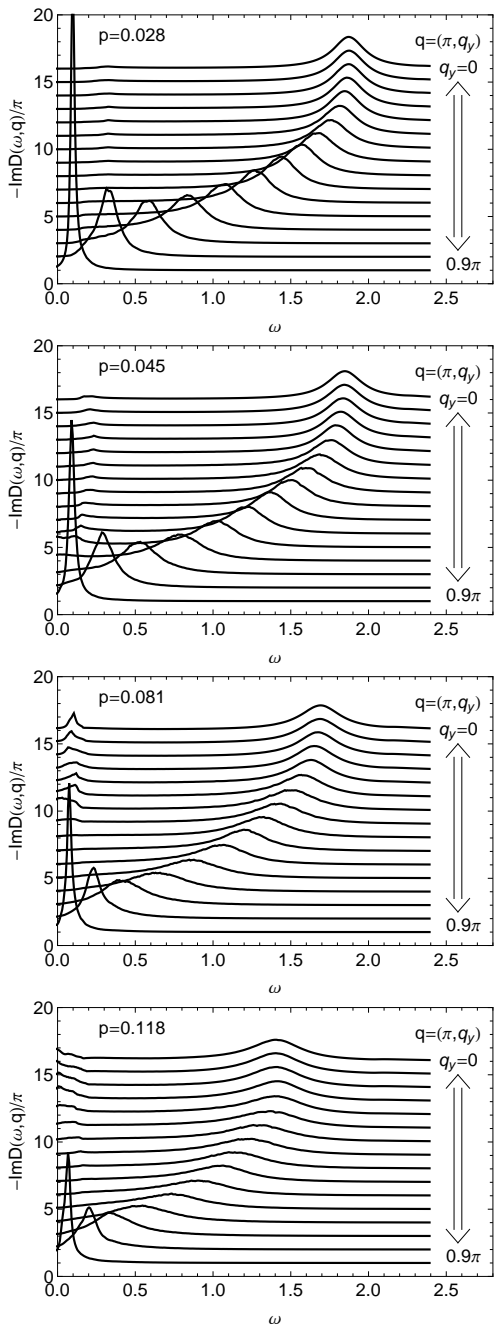}
\caption{Imaginary part of Magnon Green's function at different doping levels for the $t-J$ model with $t^{\prime}=t^{\prime\prime}=0$. } 
\label{fig:ImD_ZZ_ZA_ZB_AB_scandoping_tJ}
\end{center}
\end{figure}
Fig.~\ref{fig:ImD_ZZ_ZA_ZB_AB_scandoping} shows the imaginary part of magnon Green's function $D(\omega,{\bm q})$
at different doping levels. The low frequency behaviour, $\omega \lesssim 0.5J$  is somewhat erratic,
this is especially evident at  $p=0.094,0.142$. 
So, the calculation indicates that low energy dynamics in this approach are not selfconsistent as
we have already discussed and as we expect.
The snapshot approach does not give low energy dynamics.
However, the high energy dynamics in this approach are reliable.
Here we concentrate on the position of the topmost peak, $\omega_{\pi,0}$.
According to Fig.~\ref{fig:ImD_ZZ_ZA_ZB_AB_scandoping}
\begin{equation}
\label{om}
\omega_{\pi,0} \approx 2J(1-3.6p)
\end{equation}
As we expect from general arguments presented above there is no a $\sqrt{p}$ term in $\omega_{\pi,0}$.
Nevertheless, there is a significant softening of the zone boundary magnon with doping.

We have also performed calculations for the pure $t-J$ model ($t^{\prime}=t^{\prime\prime}=0$).
The magnon spectral function for this case is shown in Fig.~\ref{fig:ImD_ZZ_ZA_ZB_AB_scandoping_tJ}. 
In this case there is a noticeable very low energy structure practically for all momenta.
The structure is a reflection of negative compressibility of the pure $t-J$-model~\cite{sushkov04}, 
the present calculation also indicates the compressibility problem in the low energy dynamics.
Again, the high energy dynamics are reliable.
The position of the topmost peak is
\begin{equation}
\label{om1}
\omega_{\pi,0} \approx 2J(1-2.1p)
\end{equation}
In this case softening of the zone boundary magnon with doping is less pronounced.  Hence, 
the degree of magnon softening depends significantly on  values of $t^{\prime}$ and $t^{\prime\prime}$.
This is quite natural since these parameters change shape of the Fermi surface.

The result (\ref{om}) has to be compared with RIXS data. We remind that the result is theoretically
justified  only for $p < 0.15$. However, this is sufficient to claim that (\ref{om})
is inconsistent with data~\cite{LeTacon11}, provided that RIXS indeed
measures the spin response.

In particular for optimal doping ($p=0.15$) the theory predicts softening of the zone
boundary magnon by factor of 2. There is no any softening observed in experiment.
This indicates that real cuprates are correlated more strongly than the $t-t'-t''-J$ model.
We do not see a possibility to challenge physics of the parent CTI.
Therefore, in our opinion the discrepancy indicates a failure of the Zhang-Rice singlet picture
away from the heavily underdoped regime.

We would also like to comment on the validity of the single band Hubbard model for 
description of cuprates.
We reiterate again that  parent cuprates are not  Mott insulators, they are  charge 
transfer insulators~\cite{CT,Chen91}.
Therefore, one cannot directly justify the Hubbard model and usual justification is based on 
the reversed argument that if the $t-t'-t''-J$ model is valid then it must be equivalent to
some effective Hubbard model.
However, if the $t-t'-t''-J$ model and the Zhang-Rice singlet picture fails, then, in our opinion,
there is no way to justify the single band Hubbard model.
%A more technical comment concerns the three site terms~\cite{3} in the $t-J$ model derived from the Hubbard model.
%We have shown that specific values of additional hopping matrix elements $t'$ and $t''$ are important for 
%the zone boundary magnon  softening. Values of the matrix elements were found from fitting experimental hole 
%dispersion in frameworks of the $t-t'-t''-J$ model. Performing a similar analysis within the  single band Hubbard model
%one must introduce additional hopping matrix elements $t'_H$, $t''_H$ and they differ from $t'$, $t''$ due
%to the 3-cite hopping.

\section{Spin structure factor sum rule and implications for overdoped cuprates}

\subsection{The sum rule}
The spin structure factor is defined as
\begin{eqnarray}
\label{sf}
S({\bm q},\omega)=\sum_n\langle 0|{\bm S}^{\dag}_{\bm q}|n\rangle
\langle n|{\bm S}_{\bm q}|0\rangle \delta(\omega-\omega_n)\ ,
\end{eqnarray}
where ${\bm S}_{\bm q}$ is the Fourier transform of the electron spin density
\begin{equation}
\label{sd}
{\bm S}_{\bm q}=\frac{1}{N}\sum_j{\bm S}_je^{i {\bm q}\cdot{\bm r}_j} \ ,
\end{equation}
where ${\bm S}_j=\frac{1}{2}\sum_{\mu\nu}c_{j\mu}^{\dag}{\bm \sigma}_{\mu\nu}c_{j\nu}$.
Here we assume that the system is described by a model on a square lattice,
either the $t-t'-t''-J$ model (\ref{single_layer_H}), the Hubbard model, or any other {\it single band} model. 
The spin structure factor gives intensity of spin response and it has been measured in 
RIXS~\cite{Braicovich09,Braicovich10,LeTacon11,Bisogni12,Bisogni12_2,Dean12,Dean13,LeTacon13}
as well as in neutron scattering. The spin structure factor obeys the following  sum rule
\begin{eqnarray}
\label{sr}
&&\int_0^{\infty}d\omega\int_{BZ}\frac{d^2q}{(2\pi)^2} S({\bm q},\omega)=S_{rule} \\
&&S_{rule}=\frac{3}{4}\left\{(1-p)-
\frac{2}{N}\sum_j\langle 0|c_{j\downarrow}^{\dag}c_{j\uparrow}^{\dag}c_{j\downarrow}c_{j\uparrow}|0\rangle
\right\} \ .\nonumber
\end{eqnarray}
The  momentum integration here is carried over full Brillouin zone (BZ)
of the lattice.
To derive the sum rule one just use the closure relation, $\sum_n|n\rangle\langle n|=1$,
and the anticommutation relation for electron creation/annihilation operators $c_{j\sigma}$.
Note that the first term in brackets in (\ref{sr}) is just the average electron density per site,
$\frac{1}{N}\sum_{j\mu}\langle 0|c_{j\mu}^{\dag}c_{j\mu}|0\rangle=N_e/N=1-p$.

Now we discuss the sum rule in each particular model.
For noninteracting electrons
$\sum_j\langle 0|c_{j\downarrow}^{\dag}c_{j\uparrow}^{\dag}c_{j\downarrow}c_{j\uparrow}|0\rangle=
N_e^2/(4N)$. Therefore, 
\begin{equation}
\label{srn}
S_{rule}=\frac{3}{8}(1-p^2)\ .
\end{equation}
Hence, the sum rule gives 3/8 for half filling ($p=0$), and it gives 
zero for $p=1$ (no electrons) and for $p=-1$ (completely full band, two 
electrons per site).

 The $t-t'-t''-J$ model (\ref{single_layer_H}) contains a no double electron occupancy constraint
which implies that $p>0$.
In this case the second term in brackets in (\ref{sr}) is identical zero,
$\sum_j\langle 0|c_{j\downarrow}^{\dag}c_{j\uparrow}^{\dag}c_{j\downarrow}c_{j\uparrow}|0\rangle=0$,
and the sum rule reads
\begin{equation}
\label{srtJ}
S_{rule}=\frac{3}{4}(1-p)\ .
\end{equation}

For the single band Hubbard model the second term in brackets in (\ref{sr}) is nonzero, but if $U \gg t$ 
it is small.
%Here we assume that $p \geq 0$ and hence the upper Hubbard band is unpopulated.
At exactly half filling, $p=0$, the sum rule reads
\begin{equation}
\label{srHh}
S_{rule}=\frac{3}{4}\left[1-8\left(\frac{t}{U}\right)^2\right]=
\frac{3}{4}\left[1-\frac{1}{2}\left(\frac{J}{t}\right)^2\right].
\end{equation}
Here we use the standard relation $J=4t^2/U$.
For $t/J\approx 3$ the sum rule (\ref{srHh}) deviates from that for the $t-t'-t''-J$ model,
Eq. (\ref{srtJ})  only by 5\%.
The spectral weight for magnetic transitions to the upper Hubbard band, $\omega \approx U$,
is tiny, $\approx \left(\frac{t}{U}\right)^2 \approx \frac{1}{16}\left(\frac{J}{t}\right)^2$.
So, most of the weight (\ref{srHh}) is in excitation of usual magnons $\omega \leq 2J$.

Finally we present the sum rule for the Hubbard model in the dilute electron limit, $N_e/N=1-p \ll 1$.
This is the limit of normal Fermi liquid and a simple summation of ladder diagrams 
in $\sum_j\langle 0|c_{j\downarrow}^{\dag}c_{j\uparrow}^{\dag}c_{j\downarrow}c_{j\uparrow}|0\rangle$
gives the following answer for the sum rule
\begin{equation}
\label{srH1}
S_{rule}=\frac{3}{4}\left\{(1-p)-(1-p)^2\frac{2\pi^2}{ln^2\left(\frac{4.2}{1-p}\right)}\left(\frac{J}{t}\right)^2\right\} \ .
\end{equation}
The parameter $J=4t^2/U$ does not give an energy scale in this case. The spectral weight is
practically uniformly distributed over the entire band width $0 < \omega < 8t$. There is also a tiny 
weight for transition to
the upper Hubbard band, $\omega \approx U$.
In both cases, Eq.~(\ref{srHh}) and Eq.~(\ref{srH1}), the $(J/t)^2$ correction does not exceed
several per cent and hence for our purposes it can be neglected. The sum rule in the Hubbard model is very 
close to that in the $t-t'-t''-J$ model.

The RIXS 
measurements~\cite{Braicovich09,Braicovich10,LeTacon11,Bisogni12,Bisogni12_2,Dean12,Dean13,LeTacon13}
have been performed in undoped, underdoped, and overdoped cuprates,  and structure factors have been
compared between these cases. The comparison includes not only positions of energy peaks, but intensities too.
This allows us to use undoped cuprates, where theory is unambiguous, as reference point.
Therefore, in the next subsection we calculate the structure factor of undoped 2D CTI or Mott insulator
treated within the framework of the Heisenberg model.

\subsection{Spin structure factor of 2D Heisenberg model}

For the calculation we use the standard spin-wave approach briefly discussed in Section II.
The spin structure factor has been considered previously in Refs.~\cite{Igarashi92a,Stringari94}
within the  same approach, as well as by quantum Monte Carlo method in Ref.~\cite{Sandvik01}.
For our purposes we need a closed form of the structure factor over the entire BZ,
which is not addressed in the previous works, so in what follows we revisit this problem.

\begin{figure}[ht]
\begin{center}
\includegraphics[clip=true,width=0.8\columnwidth]{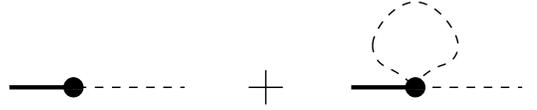}
\caption{Creation of a single magnon by the spin operator.
The solid line shows the spin and the dashed lines show magnons.
}
\label{1M}
\end{center}
\end{figure}

\begin{figure}[ht]
\begin{center}
\includegraphics[clip=true,width=0.8\columnwidth]{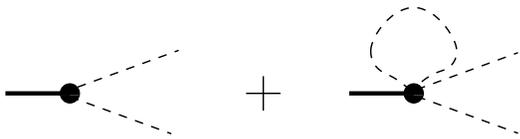}
\caption{Creation of two magnons by the spin operator.
The solid line shows the spin and the dashed lines show magnons.
}
\label{2M}
\end{center}
\end{figure}

In the calculation we keep magnon quantum corrections up to the single loop~\cite{com1}.
We remind that the staggered magnetization calculated in the single loop approximation reads~\cite{staggered}
\begin{eqnarray}
\label{st}
&&m=\frac{1}{2}-r \approx 0.304\nonumber\\
&&r=\sum_{{\bm k}\in MBZ}2v_{\bm k}^2\approx 0.196\ ,
\end{eqnarray}
where $v_{\bm k}$ is Bogoliubov parameter defined in Eqs.~(\ref{Bogoliubov_trans}),(\ref{bog}).
The intermediate state $|n\rangle$ in Eq.~(\ref{sf}) can consist of one, two, three, and more magnons.
The single magnon matrix element $\langle n|{\bm S}_{\bm q}|0\rangle \to
\frac{1}{\sqrt{2}}\langle n|S_{\bm q}^{(+)}|0\rangle +\frac{1}{\sqrt{2}}\langle n|S_{\bm q}^{(-)}|0\rangle
$ is described by diagrams shown in Fig.~\ref{1M}.
A straightforward calculation gives the following single magnon contribution to the spin structure
factor
\begin{eqnarray}
\label{1MC}
S^{(1)}({\bm q},\omega)=\frac{1}{2}[1-r]^2(u_{\bm q}+v_{\bm q})^2\delta(\omega-\omega_{\bm q}).
\end{eqnarray}
The double magnon contribution contains an elastic part determined by the staggered magnetization (\ref{st})
and inelastic part determined by the double magnon matrix element $\langle n|{\bm S}^{(z)}_{\bm q}|0\rangle$
described by diagrams shown in Fig.~\ref{2M}.
Again, a straightforward calculation gives the following two magnon contribution to the spin structure
factor
\begin{eqnarray}
\label{2MC}
&&S^{(2)}({\bm q},\omega)=[1/2-r]^2\delta(\omega)\delta({\bm q}-{\bm Q}_{AF})\\
&&+[1-r]^2\sum_{{\bm k}\in MBZ}(u_{\bm k}v_{\bm q-k}-v_{\bm k}u_{\bm q-k})^2\delta(\omega-\omega_{\bm k}-\omega_{\bm q-k})\ .
\nonumber
\end{eqnarray}
Here ${\bm Q}_{AF}=(\pi,\pi)$ is the AF vector.
Performing the $\omega$- and the ${\bm q}$-integrations in (\ref{1MC}) and (\ref{2MC}) one finds the single
and the double magnon contributions to the sum rule (\ref{sr}).
\begin{eqnarray}
\label{sr12}
&&\int_0^{\infty}d\omega\int_{BZ}\frac{d^2q}{(2\pi)^2} S^{(1)}({\bm q},\omega) =\frac{1}{2}(1-r)^2(1+2r)\approx 0.45
\nonumber\\
&&\int_0^{\infty}d\omega\int_{BZ}\frac{d^2q}{(2\pi)^2} S^{(2)}({\bm q},\omega) \\
&&=(1/2-r)^2+ r(1-r)^2(1+r)\approx 0.09+0.15=0.24 \ .\nonumber
\end{eqnarray}
In the last line of Eq.~(\ref{sr12}) we present separately contributions of the static diffraction and the 
double magnon excitation. The first line, 0.45, is slightly below the exact sum rule value 0.5.
This gives an estimate for the triple magnon contribution, $\sim 0.05$.
The estimate is consistent with accurate quantum Monte Carlo calculations~\cite{Sandvik01}.
The single and the double magnon contributions practically saturate the sum rule 
(\ref{sr}), 0.45+0.24=0.69 instead of expected 0.75.
Therefore we neglect triple magnon excitations as well as higher multiplicity excitations.
\begin{figure}[ht]
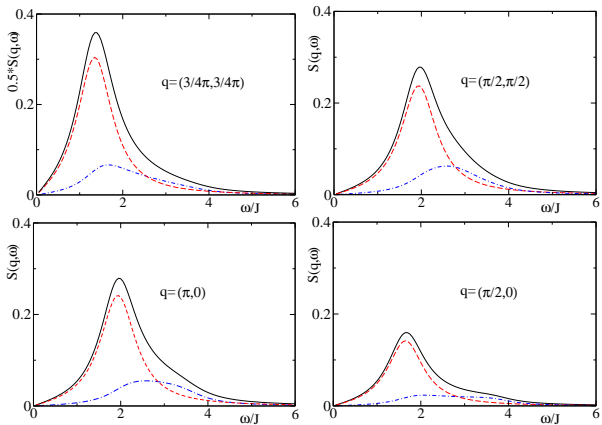

\begin{center}
\includegraphics[clip=true,width=0.45\columnwidth]{Rez075pi075pi.eps}
\includegraphics[clip=true,width=0.45\columnwidth]{Rez05pi05pi.eps}
\includegraphics[clip=true,width=0.45\columnwidth]{Rezpi0.eps}
\includegraphics[clip=true,width=0.45\columnwidth]{Rezpi20.eps}
\caption{Spin structure factor $S({\bm q},\omega)$ versus $\omega$ for 2D Heisenberg model at 
${\bm q}=(\frac{3}{4}\pi,\frac{3}{4}\pi)$, ${\bm q}=(\frac{1}{2}\pi,\frac{1}{2}\pi)$,
${\bm q}=(\pi,0)$, and ${\bm q}=(\frac{1}{2}\pi,0)$.
Solid black lines show the total structure factors, red dashed lines show 
the single magnon contributions and blue dot-dashed-lines show the double
magnon contribution. The broadening width is $\Gamma_0=J$.
Note that the first panel, ${\bm q}=(\frac{3}{4}\pi,\frac{3}{4}\pi)$, is scaled down by factor 2.
}
\label{075}
\end{center}
\end{figure}
To plot the structure factor we broaden $\delta$-functions in Eqs.~(\ref{1MC}),(\ref{2MC})
\begin{equation}
\label{br}
\delta(\omega-\Omega)\to F(\omega)=
R\frac{2}{\pi}\frac{\omega \Omega\Gamma_0}{(\omega^2-\Omega^2)^2+\omega^2\Gamma_0^2}\ .
\end{equation}
The  factor $R\sim 1$ is determined numerically from the normalization condition 
$\int_0^{\infty}F(\omega)d\omega=1$. Note that $F(\omega)$ has an  asymmetric shape with effective 
linewidth $\Gamma_{\omega}=\frac{\omega}{\Omega}\Gamma_0$.
An $\omega$ dependence of the linewidth is typical for broadening due to inelastic 
processes, for example for electric dipole transitions in atoms/molecules/nuclei 
$\Gamma_{\omega}= (\omega/\Omega)^3\Gamma_0$. 
The $\omega$-dependent broadening leads to a 
non-Lorentzian lineshape.
Calculated structure factors for $\Gamma_0=J$ and for ${\bm q}=(\frac{3}{4}\pi,\frac{3}{4}\pi)$,
${\bm q}=(\frac{1}{2}\pi,\frac{1}{2}\pi)$,
${\bm q}=(\pi,0)$, and ${\bm q}=(\frac{1}{2}\pi,0)$ are plotted in Fig.~\ref{075}.

\subsection{Spin sum rule in underdoped cuprates}

The sum rule (\ref{sr}),(\ref{srtJ}) is naturally fulfilled withing the $t-t'-t''-J$ theory of 
underdoped cuprates~\cite{milstein08}.
The small static response described by the first line in Eq.~\ref{2MC} is getting
incommensurate with doping and very quickly diminishes. It completely disappears at doping
higher than QCP at $p\approx 0.1$. Certainly this contribution does not disappear from the spin sum rule.
The corresponding spectral weight is transferred to the hourglass neck, so the parent Heisenberg model
static response becomes dynamic with the typical energy $E_{cross} \propto p^{3/2}$ 
($E_{cross}(p=0.15)= 40-50$meV). The corresponding
contribution to the spin sum rule is relatively small, $(1/2-r)^2=0.09$.
Within the chiral perturbation theory the higher energy magnetic excitations, 
$\omega > E_{cross}$ are not modified by doping besides the softening and broadening discussed in Section II.
Thus the integrated spectral intensity remains practically the same as that in the parent compound.
This prediction of the theory is perfectly consistent with RIXS data~\cite{LeTacon11}, the spin 
sum rule is naturally fulfilled because almost nothing is changed compared to the parent CTI.
A small reduction of the total spectral weight proportional to doping $p$, the right hand side in Eq.~(\ref{srtJ}),
is beyond accuracy of the chiral perturbation theory.

To avoid misunderstanding we reiterate that the $t-t'-t''-J$ model makes the
following predictions in the underdoped regime, (i) Softening of the zone boundary magnons
with doping discussed in Section II. (ii) No variation of spectral weight with doping
for high energy magnetic excitations, $\omega \gg E_{res}$.
The first prediction is inconsistent with RIXS data while the second one is 
perfectly consistent with the data.

\subsection{Spin sum rule in overdoped cuprates}
Even if the $t-t'-t''-J$ model was valid for overdopped cuprates there is no any
controlled
theoretical technique to analyze the model in this regime. As we have already explained the chiral 
perturbation theory can at most be extended up to optimal doping.
Independently of underlying microscopic model there are numerous experimental indications
that overdoped cuprates behave like ordinary Fermi liquids,
See Ref. \onlinecite{LeTacon13} for summary of the indications.
In this subsection we analyse how the exact spin sum rule (\ref{sr}) is saturated in highly overdoped 
Tl$_2$Ba$_2$CuO$_{6+\delta}$ with doping $p\approx 0.27$. 
We do not have a microscopic theory for this case. However, whatever is the model/theory 
the total integrated spin spectral weight must be equal to 
$0.75(1-0.27)\approx 0.55$, see Eqs. (\ref{sr}), (\ref{srtJ}), (\ref{srHh}), (\ref{srH1}).

\begin{figure}[ht]
\includegraphics[clip=true,width=0.85\columnwidth]{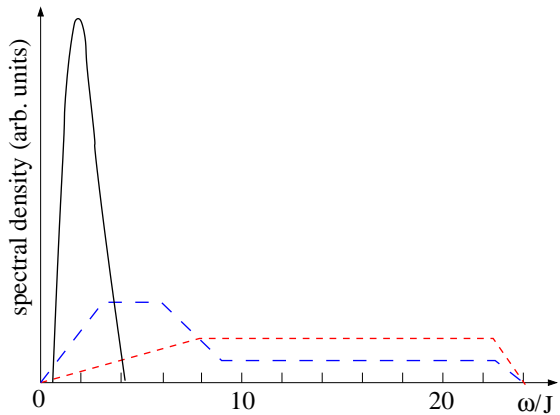}
\caption{A schematic picture of q-integrated spin spectral density.
The solid black line follows from RIXS data~\cite{LeTacon13},
so this is the measured spectral density.
The red short-dashed line corresponds to a noninteracting NFL,
the band width is $8t\approx 24J\approx 3.4$eV. The blue long-dashed line is a sketch
for a strongly interacting NFL.
Note that the areas over all the curves is the same.
 }
\label{sp}
\end{figure}

The message of Ref.~\onlinecite{LeTacon13} is that while the low energy, $\omega < 100$meV, spin response
is almost diminished in Tl$_2$Ba$_2$CuO$_{6+\delta}$, the high energy spin response, $\omega > 100$meV
is the same as that in the parent compound. The positions of spectral maximums 
and the spectral weights are the same~\cite{com2}. 
We have a reliable theory for the parent compound and hence we can calculate 
contribution to the sum rule from $\omega > 100$meV. Momentum and $\omega >100$meV integration of 
Eq.~(\ref{1MC}) gives 0.37, and similar integration of Eq.~(\ref{2MC}) gives 0.14. 
Hence, the measured spectral weight $0.37+0.14=0.51$ is close to 0.55 expected from the
exact sum rule.
Thus, we conclude that the {\it entire} spin spectral weight in heavily overdoped
Tl$_2$Ba$_2$CuO$_{6+\delta}$ is located in the same energy range $\sim 2J+broadening$ as that
in undoped CTI. 
In fully uncorrelated NFL the spectral weight is almost uniformly distributed over the entire bandwidth,
$\Delta E \approx 8t\approx 24J\approx 3.4$eV. A schematic picture of the q-integrated spin spectral density is
shown in Fig.~\ref{sp}.
The solid black line has been extracted from RIXS data~\cite{LeTacon13}
using the parent compound normalization as it has been described above, 
so this is the measured spectral density.
The red short-dashed line corresponds to a fully uncorrelated NFL.
The blue long-dashed line is a cartoon for correlated NFL, 
the effective bandwidth is reduced by 3 times compared to the noninteracting NFL
to imitate the effective mass measured in magnetic oscillations~\cite{Vignolle08}.
Obviously the measured spectral density is dramatically different from what one can expect
from a NFL model.  This analysis supports the statement of the experimental paper~\cite{LeTacon13}
that the observed magnetic response is inconsistent with the NFL picture in 
highly overdoped Tl$_2$Ba$_2$CuO$_{6+\delta}$.

%We would like also to note that the observed very low energy concentration of the spin response
%indicates a spin charge separation~\cite{Holt12}.
%While the spin response is concentrated at the low energy, the solid curve in Fig.~\ref{sp},
%the charge response for sure is concentrated at the high energy, like dashed or dashes-dotted lines
%in Fig.~\ref{sp}.

\section{Conclusion}
In summary, stimulated by recent RIXS data we analyze underdoped and overdoped regimes
of cuprates. (i) Our analysis of the underdoped regime is based on the 
$t-t^{\prime}-t^{\prime\prime}-J$ model. The model is treated within the controlled
chiral perturbation theory with doping being the small expansion parameter.
Our calculation demonstrates a significant softening of the high energy ($\omega > 100$meV)
magnetic response with doping, see Eq. (\ref{om}).
This is inconsistent with RIXS data which show that the high energy magnetic response
is practically doping independent.
(ii) Our analysis of the heavily overdoped regime is based on the exact spin sum rule.
We demonstrate that the observed in RIXS magnetic response saturates the spin sum rule.
This implies that the entire momentum integrated magnetic response is concentrated
in the energy interval $\sim2J$ similar to that in the undoped compound.
Such energy concentration of the magnetic response is not consistent with a
normal Fermi liquid model.

In our opinion the discussed inconsistencies most likely indicate a failure
of the Zhang-Rice singlet picture away from the underdoped regime.
According to RIXS measurements antiferromagnetic correlations are stronger than 
that predicted by the 
$t-t^{\prime}-t^{\prime\prime}-J$ model. The analysis suggests that electron 
spins on copper sites are practically not influenced by doping besides of
the relatively low frequency, $\omega \ll 100$meV, fluctuations.
Further investigation is certainly needed to address the indicated problems.

We thank G. Khaliullin, M. Le Tacon, B. Keimer, T. Tohyama, M. Berciu, and G. Sawatzky for 
stimulating discussions.


\begin{thebibliography}{99}
\bibitem{com2} One can raise some doubts if the observed low energy response~\cite{LeTacon13}
has a fully magnetic origin. However, a nonmagnetic low energy response would be even more puzzling
than the magnetic one.
\end{thebibliography}
\end{document}